# Low-cost Relevance Generation and Evaluation Metrics for Entity Resolution in AI

**Venkata sai Varada**  **Mina Ghashami**  **Jitesh Mehta**  **Haotian Jiang**  **Kurtis Voris**
Alexa Entertainment, Amazon
{vnk, ghashami, mjitesh, haotij, kurtv}@amazon.com

## Abstract

Entity Resolution (*ER*) in voice assistants is a prime component during run time that resolves entities in user's request to real world entities. ER involves two major functionalities 1. *Relevance generation* and 2. *Ranking* . In this paper we propose a low cost relevance generation framework by generating features using customer implicit and explicit feedback signals. The generated relevance datasets can serve as test sets to measure *ER* performance. We also introduce a set of metrics that accurately measures the performance of *ER* systems in various dimensions. They provide great interpretability to deep dive and identifying root cause of *ER* issues, whether the problem is in relevance generation or ranking.

## 1 Introduction

In smart home devices with multiple modalities of input, e.g: text and voice, such as Fire TV, Chromecast, Apple TV powered by voice assistants like Alexa, Google Assistant, Siri, *ER* is a process of resolving entities in users' query to actual entities from a predefined catalog. Voice assistants (Hoy, 2018) comprise of Automatic Speech Recognition (*ASR*) which transcribes speech to text and is then passed to Natural Language Understanding (*NLU*) module. *NLU* predicts the intent of the user utterance and the slotting and hands it over to *ER* . For example:

- *ASR* Recognition : *play bridgerton on Netflix*
- *NLU* Intent: *PlayVideoIntent*
- *NLU* slotting: *VideoName: bridgerton, AppName: Netflix*

*ER* then performs search requests in a data catalog and provides a list of possible entities relevant to the user's request.

- *ER* output: *[{Bridgerton, ID: 1234, confidence: High}, {Bridgetown, ID: 5678, confidence: Medium} . . .]*

*ER* systems output k (normally 5) entities that are most likely to be relevant with a confidence $bin \in$[low, medium, high] for each output entity. The correct entity expected by the user is referred as relevance. An Overview of ER is given in section 2.

Evaluation of *ER* systems is of prime importance as resolving to a wrong entity could lead to user friction, for example, playing a wrong movie/song. These systems need to be fail proof and have very low margin for error. For evaluating *ER* systems one needs ground truth relevance test set and metrics that provide the true performance of *ER* systems. With respect to test sets, human annotated data labeling is a rich source of relevance, but is expensive and time consuming. Hence there is a need to generate relevance in a low cost fashion. In this work, we use implicit customer feedback collected from other modalities of the system and explicit feedback from popular IMDb (Wikipedia, 2021a) data source to generate relevance in section 3. With respect to evaluation metrics, ER typically borrows traditional information retrieval metrics including Precision and Recall (Wikipedia, 2021b). Recall is the fraction of the relevant documents to the query that are successfully retrieved. Precision is the fraction of the documents retrieved that are relevant to the user's request.

$$Recall = \frac{\{relevant\ documents\} \bigcap \{retrieved\ documents\}}{\{relevant\ documents\}} \quad (1)$$

$$Precision = \frac{\{relevant\ documents\} \bigcap \{retrieved\ documents\}}{\{retrieved\ documents\}} \quad (2)$$

The {retrieved documents} denotes the set of documents given by information retrieval system which include both relevance generation and ranking components.

A slight modification to these metrics is to compute precision@k and recall@k by calculating precision and recall only on a subset of top k results. e.g., Precision@10 corresponds to the number of relevant results among the top 10 retrieved documents. Using pairwise Precision, Recall and F1 scores (Menestrina et al., 2010) and cluster level Precision, Recall metrics (Barnes, 2015) is another popular approach when performing *ER* across databases. All the existing metrics focus only on the top k output items with out considering the confidence with which they are returned by the system. More often than not, the true relevant item has Medium or Low confidence which is not returned to the user making it as a performance gap in ER without knowing if the problem is in relevance generation or in ranking. Hence, there is a need to design metrics to address aforementioned challenge. We discuss a new set of metrics in section 4.

Our contributions from this paper include

- A low cost and efficient way to automatically generate relevance test sets.
- A set of novel evaluation metrics for ER systems providing deep insight into the root cause of failure, aiming to improve customer satisfaction.

## 2 Overview of Entity Resolution

*ER* systems follow the classic two-stage information retrieval dichotomy constituted by a *relevance generation* component and a *ranking* component. All the documents related to the application are indexed in a search database so that it is easy to find and retrieve them later. Upon receiving a user's query, the *relevance generation* component queries it against the search module of *ER* to retrieve relevant documents. This component uses various searching and text matching techniques to find the near perfect entities.

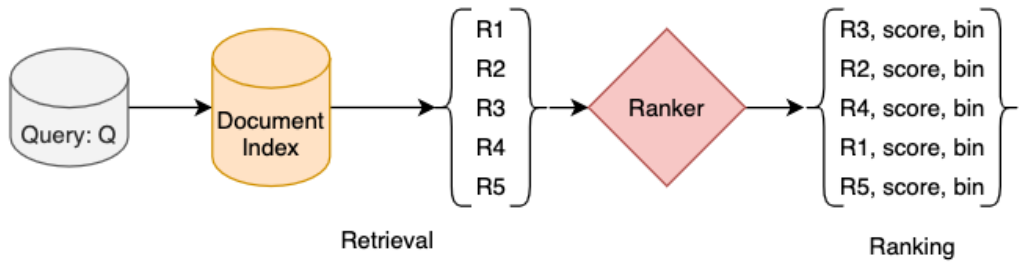

Figure 1: Overview of Entity Resolution

The retrieved entities are then passed to the *ranker* component which ranks them and additionally provide a confidence score and a confidence bin. The scores and bin assignment reflect how relevant the retrieved results are to the input query. While the scores are bounded float numbers, the confidence bins can be one of {high, medium, low}. These bins are assigned based on the score computed by the *ranker* which tells how confident the ranker is. Figure 1 gives an overview of *ER* .

## 3 Relevance generation for Evaluation sets

Relevance in search is defined as the measure of accuracy between the search query and the resultant entity. Constructing relevance for query-entity association is a crucial aspect in *ER* . There is a lot of work being done in designing and establishing a human in the loop annotation workflow to more objectively make judgement on query-entity relevance. These workflows are often carried out by internal annotators or by crowd sourcing. Either of these approaches incur a heavy cost and more importantly includes exposing not only user's request but also full context of user's request and other related information to annotators to accurately judge the relevance and correct it incase of poor relevance.

### 3.1 Feature Generation for *ER*

In this section, we discuss how we constructed highly reliable features that help us in inferring relevancy through out *ER* . To do so we used two main datasets *Internet Movie Database (IMDb)* and *Search-and-click*. Below, we explain these datasets and how we constructed the relevant features from them.

#### 3.1.1 Importance Score Feature

Internet Movie Database (IMDb) (Wikipedia, 2021a) is an online database of information related to films, television programs, videos, and streaming content online. It not only provides information related to cast, plot summaries, personal biographies, but also provides valuable user feedback including ratings, critical reviews etc. Additionally, IMDbPro offers a ranking (IMDb) for every title. It uses proprietary algorithms that take into account several measures of popularity. The primary measure is how many and what content people are looking at on IMDb. Figure 2 explains how IMDb signals are processed.

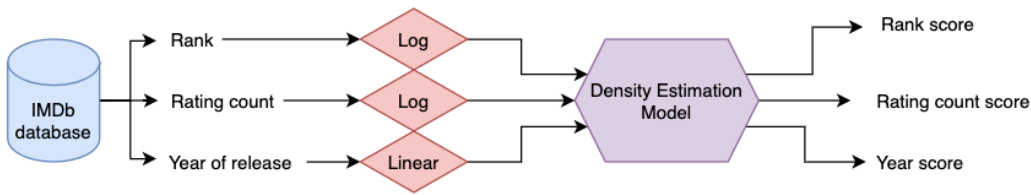

Figure 2: Overview of IMDb data

We primarily use three features from IMDb dataset to devise an *importance score* which captures the importance of a video. The three features we used are: 1)*release year*; the year in which the title is released, 2)*rank*; ranking of the title, and 3)*rating count*; the number of ratings received for the title. As these features are continuous and unlabelled, we applied density estimation models to them. Log-based density estimation models are applied for *rank* and *rating count* to better out stand the difference between lower and higher ranks and rating counts, and a linear model is applied to *release year* as recency is naturally a linear notion.

After running the above three IMDb features through density estimation models, and obtaining the corresponding scores of *release year score, rank score,* and *rating count score*, we compute the final *importance score* as:

$$\text{importance score} = f(\text{release-year-score}, \text{rank-score}, \text{rating-count-score}) \quad (3)$$

Where $f$ is a linear function.

### 3.1.2 Click Through Rate Feature

Search requests such as *"find spider man movies"*, and *"search for cocomelon"* are common type of queries on multi-modal devices, furthermore, the most significant type of voice and text initiated queries are title-only searches such as *"cocomelon"*, i.e. customers ask for just titles with out verbs including play or search. These queries along with the corresponding results shown to customers are captured in the *search-and-click dataset*. This dataset is an internal data that is processed and anonymized to de-identify any user related information. Specifically, this dataset contains search queries through voice or text, the impressions shown, and the corresponding click information. This is a rich source of customers' implicit feedback, and so we use it to compute *Click Through Rate (CTR)* measure.

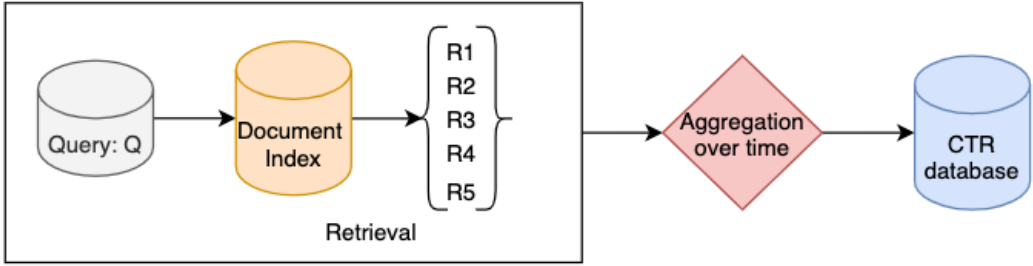

Figure 3: Overview of $CTR$ aggregation

Consider $q$ to be a search query, and let $\{r_1, r_2 \cdots r_n\}$ denote the set of corresponding retrieved results from a search engine. For any result $r_i$, we define $nimp(q, r_i)$ as number of impressions of $r_i$ with respect to query $q$, and $nclick(q, r_i)$ as number of clicks on $r_i$ with respect to query $q$. Then $CTR$ of pair $(q, r_i)$ is calculated as following:

$$CTR_{(q,r_i)} = \frac{nclick(q, r_i)}{nimp(q, r_i)} \qquad (4)$$

Figure 3 illustrates how $CTR$ is aggregated. *Search-and-click dataset* provides an important and a reliable signal as users choose which entity among search results is the correct entity to their query.

### 3.2 Filtering Logic for Datasets:

Since $CTR$ data is a pair wise dataset, It contains a lot of noise. Though we may have a similar number of aggregated impressions for a query across different results, we may see a very skewed distribution of clicks across the results because in most of the cases there is an obvious result which will have a higher number of clicks. In order to eliminate noise, we apply filters on the $CTR$ dataset to include high quality entities. For example, We do not want to include the cases with just say 4 impressions and 3 clicks as it leads to a 0.75 $CTR$ so a filter on impressions is necessary to avoid numeric bias in the denominator for $CTR$. Additionally one needs to apply a filter on the final $CTR$ to include high quality pairs. These filters can be chosen based on a trade off between quantity and quality of relevance items needed. Similarly a filter is applied on the importance score to include important videos.

We make use of these two data sources and perform merge logic on entities to extract relevant entities. The generated relevant entities dataset is a rich source to be used as test sets to evaluate and measure the relevance of search engines. In order to measure the quality of our relevance generation, we extracted a sample of 2000 data points from the relevance generation set and compared it against an expert human annotation. We achieved 95% accuracy.

Additional to the discussion in 3, the problems with human annotation further include high cost and time, scalability, additional time involved in training new associates, maintaining the ever changing annotation guidelines and furthermore it may not be 100% accurate considering the human error. The method of automatically generating relevance test sets using implicit customer feedback and alternative data sources like IMDb and with high 95% accuracy has proven to be very cost and time effective for us.

## 4 Novel Evaluation Metrics

There are instrumented metrics in voice assistants that measure success of users' utterances, however they are not suited to evaluate performance of an *ER* system. In the sections below, we first explain what existing metrics are and why they can not measure performance of *ER* . Second, we propose our novel evaluation metrics and demonstrate our experimental results.

### 4.1 The Problem with Existing Metrics

The existing metrics in voice assistants measure success of users' utterances through their interactions with the device. These measures take into account the users' requests and the domain of operation. For instance, if the request is to *"turn on living room lights"*, a successful outcome is for the device to turn on living room's lights. In music domain, if user requests the device to play a song, a successful outcome is for the device to playback the correct song; this notion of *correctness* is measured by the number of seconds user listens to the song. The success threshold on number of seconds varies by domain. For instance, in video domain, where requests are of form *"play ⟨ video-name ⟩"*, the threshold is higher than in music domain. If an interruption, either by voice or remote, happens within the defined threshold, it is considered a failure playback. These kind of success measuring metrics are called *Playback Success Metrics (PSM )*.

Note that an *ER* system bridges the gap between *NLU* which interprets users queries, and the speechlets which deliver services to the users. Relying on speechlets' outcomes and their corresponding metrics such as *PSM* to measure *ER* 's performance, results in crude approximations as they heavily undermine true performance of an *ER* . To give an intuition on the extent of that, notice that for a user query even if the *ER* system retrieves the correct entity, the playback experience may still end in failure for various reasons including but not limited to: 1) the particular app not being installed or not enabled on the device (e.g. Spotify for music domain, Netflix for video domain), 2) user not entitled/subscribed to the service (e.g. user has not entered his credentials for Hulu service), 3) errors in *ASR* (e.g. misunderstanding *"Bridge and Ton"* for *"Briderton"*), and 4) errors in *NLU* (e.g.*"christmas tree mckenzie"* for *"christmas tree"*, where *mckenzie* was not intended for Alexa).

Furthermore, *PSM* is an online metric; that means production level codes need to be implemented and experiments have to run for several weeks to observe and analyze the impact of the experiments. This dependency of online metric makes it a costly and time consuming affair. This marks the necessity to have curated metrics that can measure true performance of *ER* systems.

### 4.2 Our Proposed Metrics

To overcome the limitations of online *PSM* along with limitations in existing offline metrics discussed in section 1, such as not considering the confidence bin of the ranker, not fully interpretable which limits us to know the root cause of failure whether the problem is in retrieval or ranking module, we propose a set of new evaluation metrics

$$Recall@k@bin = \frac{\{relevant\ documents\} \bigcap \{top\ k\ retrieved\ documents\ with\ bin\}}{\{relevant\ documents\}} \quad (5)$$

$$Precision@k@bin = \frac{\{relevant\ documents\} \bigcap \{top\ k\ retrieved\ documents\ with\ bin\}}{\{top\ k\ retrieved\ documents\ with\ bin\}} \quad (6)$$

Where $k$ is an integer typically less than 5, and $bin \in$ [low, medium, high] refers to the confidence bin. The {retrieved documents} denotes the set of retrieved documents by *ER* system which include both relevance generation and ranking components.

These metrics provide information about the performance of *ER* system at various granularity levels. For instance, *Recall@5 high* and *Precision@5 high* enable us to understand the true performance of *ER* . These metrics along with *Recall@k* and *Precison@k* will tell us if the problem is in retrieving or in ranking. For instance, if *Recall@k* and *Precison@k* are lower the problems is with the *relevance generation* component. If *Recall@k* is high and *Recall@k@high* is low, it implies that the *relevance generation* component was able to retrieve the right entity, but the ranker was not confident enough to provide a high score and there by a high bin. Additionally *Precision@k@medium* and *Recall@k@medium* metrics reveal the performance of *ER* with respect to the medium bin; explaining if the problem is with binning, If *@medium* and/or *@low* metrics are high, we could adjust the binning logic to improve *@k* high metrics.

We have performed various experiments with these metrics aiming to improve *ER* . Table 1 shows how the novel metrics are better suited to track the performance as compared to *PSM* .

Table 1: Novel metrics showing improvement for experiments while Playback Success is has minimal effect ⟨ k =5 ⟩, "-" indicates degration and "+" indicates improvement.

| Experiment | Recall@k@High | Precision@k@High | Precision@1@High | PSM |
|---|---|---|---|---|
| **Experiment** 1 | -0.89% | +16.34% | +16.91% | -3.75% |
| **Experiment** 2 | -0.44% | +30.71% | +33.33% | +0.42% |
| **Experiment** 3 | +1.77% | +34.04% | +38.29% | -0.42% |
| **Experiment** 4 | +10.21% | +55.68% | +61.97% | 3.33% |

As we can see from the *Experiment 4* row, despite all the experiments aiming to improve the system, *PSM* improved only 3.33%, whereas the *Recall@k@High* show 10% improvement, *Precision@k@High* metrics show 55% improvement and *Precision@1@High* metrics show 61% improvement. Note that these experiments are not disjoint and later experiments include the benefit of previous experiments.

## 5 Conclusion and Future work

In this paper, we present a novel low cost mechanism to generate relevance using implicit signals from search and click database by computing customer feedback (CTR) and by using popularity signals such as rank, rating count, release year through computing weighted importance score from open source databases like IMDb. Being the bridge between NLU and text to speech systems, ER systems are underestimated when measured by playback success metrics and traditional information retrieval metrics have some missing pieces, we show case a set of novel metrics that not only accurately measure the ER systems but also takes into account the details like confidence binning. These metrics also help us to know if the problem is in retrieval or in the ranking providing a great level of granularity to interpret and improve ER systems. As a next step of improving the relevance generation, we plan to include additional features like other modalities and signals, information from wikipedia, constructing a domain specific knowledge graph to identify relevant entities.